\documentclass[lettersize,journal]{IEEEtran}

\usepackage{amsmath,amssymb,amsthm,amsfonts}
\usepackage{algorithmic,algorithm,nicefrac,array}
\usepackage[caption=false,font=normalsize,labelfont=sf,textfont=sf]{subfig}
\usepackage{textcomp,stfloats,url,verbatim}
\usepackage{graphicx,xcolor,tikz}
\usepackage{cite}
\usepackage{color}
\usepackage{soul}
\setstcolor{red}
\usetikzlibrary{patterns}
\usetikzlibrary{arrows.meta,arrows}

\newtheorem{remark}{Remark}[section]

\newcommand{\winlength}{\theta}
\newcommand{\standardwinlength}{L}
\newcommand{\support}{L}
 
\newcommand{\complex}{j}

\newcommand{\col}{n}

\newcommand{\f}{m} 
\newcommand{\tap}{\omega}

\newcommand{\loss}{\mathcal{L}}
\newcommand{\ent}{\mathcal{E}}

\newcommand{\spec}{\mathcal{S}}

\newcommand{\dR}{\mathbb {R}}
\newcommand{\dZ}{\mathbb {Z}}
\newcommand{\dN}{\mathbb {N}}
\definecolor{darkred}{rgb}{.82,0,0}
\definecolor{BleuBleu}{RGB}{0 50 150}
\definecolor{darkgreen}{rgb}{0,.4,0}
\definecolor{lmcolor}{rgb}{0,.6,0}
\definecolor{ymcolor}{rgb}{1,.4,0}
\definecolor{abcolor}{rgb}{0,.6,.9}
\definecolor{mlcolor}{rgb}{0.5,.2,.5}

\usepackage{xcolor}
\usepackage[colorlinks=true,linkcolor=red,citecolor=blue,urlcolor=red]{hyperref}

\begin{document}

\title{
Learnable Adaptive Time-Frequency Representation via Differentiable Short-Time Fourier Transform 
}

\author{M. Leiber, 
Y. Marnissi, 
A. Barrau, 
S. Meignen,
and L. Massoulié
}

\maketitle

\begin{abstract}

The short-time Fourier transform (STFT) is widely used for analyzing non-stationary signals. However, its performance is highly sensitive to its parameters, and manual or heuristic tuning often yields suboptimal results. To overcome this limitation, we propose a unified differentiable formulation of the STFT that enables gradient-based optimization of its parameters. This approach addresses the limitations of traditional STFT parameter tuning methods, which often rely on computationally intensive discrete searches. It enables fine-tuning of the time-frequency representation (TFR) based on any desired criterion. Moreover, our approach integrates seamlessly with neural networks, allowing joint optimization of the STFT parameters and network weights. The efficacy of the proposed differentiable STFT in enhancing TFRs and improving performance in downstream tasks is demonstrated through experiments on both simulated and real-world data.


\end{abstract}


\section{Introduction}
\label{sec:1_intro}
Fourier theory is a cornerstone of signal processing and finds widespread application across science and engineering. The \emph{short-time Fourier transform} (STFT) is a fundamental technique for analyzing non-stationary signals in diverse fields, including audio processing \cite{stafford1998long}, medicine diagnostics \cite{huang2019ecg}, and vibration analysis \cite{leclere2016multi}. STFT-based representations such as spectrograms, mel spectrograms, and the constant-Q transform, are established tools for visualizing, understanding, and processing non-stationary signals. In recent years, the integration of \emph{time-frequency} (TF) analysis with machine learning has garnered significant attention, with STFT-based representations being extensively used in tasks such as speech recognition \cite{gong2021ast}, speech enhancement \cite{nortier2024unsupervised}, music detection \cite{schluter2014improved}, data augmentation \cite{park2019specaugment}, and source separation \cite{defossez2021hybrid}, among others. The combination of neural networks and spectrograms has demonstrated remarkable success in these applications, leveraging the ability of neural networks to learn intricate patterns and the capacity of the spectrogram to capture essential signal characteristics in the TF domain.

However, the performance of these techniques is critically dependent on the \textit{analysis window} (type and length) employed in the STFT definition, as well as the TF grid determined by the hop length. These parameters significantly influence the accuracy and interpretability of the resulting representation \cite{rabiner2007introduction, garcia2022role, barai2023empirical, deng2014foundations}. Common window functions include Hann, Hamming, and Gaussian windows, with the optimal choice often being application-specific \cite{prabhu2014window}. The \textit{window length} determines the trade-off between temporal and frequency resolution, in accordance with the Heisenberg uncertainty principle \cite{havin2012uncertainty}. The \textit{hop or overlap length}, which defines the shift between successive analysis windows, controls the balance between temporal resolution (i.e., capturing the smooth evolution of frequency content) and computational cost, as well as the temporal positioning of the analysis frames. 

This paper proposes a novel gradient-based optimization approach for tuning STFT parameters, building upon our prior works \cite{dstft, dastft, dstft2}. Gradient-based optimization has attracted considerable interest in the scientific community due to increasing computational resources and the remarkable success of neural networks trained via backpropagation \cite{rojas1996backpropagation, wythoff1993backpropagation, lecun2015deep}. The unified framework we introduce here is particularly simple to implement and offers the advantage of seamless integration with neural networks, where the STFT can be viewed as a network layer with STFT parameters acting as weights. Furthermore, it provides enhanced computational efficiency compared to traditional techniques like adaptive STFT \cite{astft} or variable STFT \cite{lee2015variable}, which typically involve optimizing over a discrete parameter set using computationally intensive greedy algorithms. Another key advantage of our optimization framework is that, by operating on real-valued parameters, it can achieve more accurate optimal values compared to methods restricted to a discrete set of parameter candidates \cite{kwok2000improved,Zhong}. 
 
A key contribution of this work is the demonstration that the STFT is differentiable with respect to its window and hop lengths, provided these parameters are treated as real-valued. This differentiability enables the computation of optimal parameters using gradient-based optimization via forward and backward propagation. The remainder of this paper is organized as follows: Sec. \ref{sec:2_stft} provides the necessary background on the STFT, and Sec. \ref{sec:rel_works} reviews related work on parameter adaptation. In Sec. \ref{sec:3_dstft}, we introduce our differentiable STFT formulation with respect to the window and hop lengths. Sec. \ref{sec:4_grad} presents the partial derivatives and backpropagation formulas. Sec. \ref{sec:computation} details the numerical implementation and computational complexity. We then propose two approaches for optimizing the STFT parameters through applications: a representation-driven optimization approach in Sec. \ref{sec:5_exp} and a task-driven optimization approach in Sec. \ref{sec:task_driven}. Finally, we provide in Sec. \ref{sec:6_ccl} some concluding remarks.

\section{Background and related works}
\label{sec:2_}
\subsection{Short-Time Fourier Transform}
\label{sec:2_stft}

The STFT operates in two stages: first, the input discrete-time signal is segmented into overlapping frames using a window function; second, the \emph{discrete Fourier Transform} (DFT) is computed for each frame. This process yields a two-dimensional complex-valued matrix, $\spec_{\tap_\standardwinlength}[\col, \f]$, where $\col$ and $\f$ are integer indices standing for the time frame and frequency bin, respectively. More formally, the STFT involves sliding a window function, or \emph{tapering function}, of even length $L$, across the input signal $s$. At each time frame, centered at a temporal position $t_\col$, the DFT of the windowed signal segment is computed to analyze the local frequency content.
Let $\tap_L$ denote the tapering function, the STFT of the signal $s$ can be expressed as:
\begin{equation}
\label{eq::eq1}
\spec_{\tap_\standardwinlength}[\col,\f] = \sum_{k=t_\col - \standardwinlength/2}^{t_\col+\standardwinlength/2-1}\tap_\standardwinlength[k-t_\col] s[k] e^{\frac{-\complex2\pi k \f}{\standardwinlength}},
\end{equation}
where $j$ is the imaginary unit. The integer time indices $t_\col$, referred to as the \emph{temporal positions}, are typically equally spaced. In such cases, the spacing between consecutive temporal positions, $ H = t_{\col}-t_{\col-1}$, is constant and is known as the \emph{hop length}.
The hop length is often defined as a fraction of the window length $\standardwinlength$ using an overlap ratio $0<\alpha<1$: $H = \lfloor \alpha \standardwinlength \rfloor$ where $\lfloor \cdot \rfloor$ denotes the floor operation. It is important to note that in Eq. \eqref{eq::eq1}, $t_\col$ represents the center of the window applied to the signal. Different temporal shifts of the window can be achieved by varying $t_\col$ over discrete values.

\subsection{Related Works}
\label{sec:rel_works}
This section provides an overview of existing approaches aimed at enhancing the TFR obtained from the STFT, with a particular emphasis on methods that strive to achieve higher energy concentration in the TF plane. These approaches can be broadly classified into two main categories: pre-processing methods, which focus on determining optimal STFT parameters, and post-processing techniques, which involve applying transformations to an initially computed STFT.

\subsubsection{STFT parameters optimization}
\label{sec::param_opt}

The concept of optimizing STFT parameters, particularly the window length, has been explored in numerous studies. Techniques such as \emph{adaptive STFT} (ASTFT) \cite{astft} and variable STFTs (VSTFT) \cite{lee2015variable} were developed to address the inherent trade-offs associated with fixed window lengths.

Various implementations of ASTFT exist. In \cite{Zhong}, a method for adapting the window length along the time axis was proposed; however, this approach is less effective for multi-component signals. Adaptations of the window length in the TF plane have also been introduced, based on local chirp-rate estimation\cite{Pei}, or by minimizing a correlation-based criterion at each TF point \cite{kwok2000improved}.    
These methods typically employ a variable integer window length to achieve a finer control over the temporal and frequency resolution trade-off based on local signal characteristics or the resulting transform. However, determining the optimal parameters often involves computationally expensive discrete optimization techniques, such as grid search or trial-and-error over a predefined set of values. 
For instance, in TF window adaptation-based ASTFT, finding the optimal parameters using grid search can become rapidly prohibitive, since grid search which exhaustively evaluates all combinations of parameters, incurs an exponential computational cost of $O(P^n)$, where $P$ is the parameter space size (e.g., number of possible window lengths) and $n$ is the number of independent TF regions. 

In VSTFT \cite{lee2015variable}, the window length is varied based on the local instantaneous frequency, estimated within each window slice over time, with the length being inversely proportional to the instantaneous frequency. However, this local estimation can be unreliable in the presence of multi-component signals. An extension of the Gabor transform, allowing for frequency (resp. time) resolution that changes over time (resp. frequency), was proposed in \cite{balazs2011theory}. However, this method does not consider simultaneous adaptation in both time and frequency.

In contrast to these traditional methods, which have primarily optimized the window size using discrete searches over a limited set of candidates, recent research has focused on optimizing STFT parameters via gradient descent. This involves treating STFT parameters as continuous, real-valued variables. For example, a gradient-based optimization of the window length was proposed in \cite{dstft}. This idea was extended in \cite{dastft} to include adaptive window lengths varying in both time and frequency. The optimization of the hop length using gradient descent was introduced in \cite{dstft2}. In \cite{Zhao}, STFT parameters were optimized for sound classification, specifically using a Gaussian window. \cite{marx2023differentiable} and \cite{he2023idsn} have explored the optimization of window length for bearing fault diagnosis.

\subsubsection{Post-processing techniques}

Numerous post-processing techniques have been explored to enhance the readability of spectrograms after their initial computation. These include reassignment methods \cite{kodera1976new, kodera1978analysis, Auger, Flandrin, auger1995improving} and synchrosqueezing transforms \cite{auger2013time, thakur2011synchrosqueezing, oberlin2014fourier, thakur2013synchrosqueezing, oberlin2015second, behera2018theoretical, li2020adaptive, fourer2016recursive}. 

Synchrosqueezing techniques aim to improve the sharpness of the TFR by reassigning the STFT coefficients of a signal. They offer an advantage over classical reassignment methods \cite{auger1995improving} by preserving the invertibility of the transform. However, the effectiveness of the reassignment process in synchrosqueezing, particularly when dealing with multi-component signals, heavily relies on the degree of separation between the modes in the TF plane. If two modes are closely spaced, and if the analysis window is not appropriately chosen, the reassignment may not be effective. To address this, new versions of the synchrosqueezing transform have been proposed that incorporate local time adaptation of the window \cite{li2020adaptive2}. Unfortunately, these transforms still do not allow for simultaneous adaptation of the window in both time and frequency. Another significant limitation of such reassignment processes is their sensitivity to noise. Furthermore, 
the optimization was based on a grid search. 
Synchrosqueezing was also adapted to the short-time fractional Fourier transform in \cite{shi2020novel}, and an improvement including window adaptation was proposed in \cite{zhao2023synchrosqueezing}. In that paper, the fractional order was also optimized locally only in time simultaneously with the window length, and 
grid search was considered for optimization. 
Finally, it is important to note that such an adaptation of the short-time fractional Fourier transform is essentially used , and very useful,  to deal with multicomponent signals containing parallel modes.

\subsubsection{Discussion}

It is worth noting that post-processing techniques share a common goal with the STFT parameter optimization methods:  to enhance the readability and interpretability of TFRs. However, their approaches differ fundamentally. Specifically, while ASTFT and VSTFT aim to adapt the TF resolution locally during the STFT computation, methods like reassignment and synchrosqueezing are applied to a STFT computed with fixed parameters, potentially optimized beforehand. As previously mentioned, the success of reassignment processes is contingent upon the initial STFT being well-suited for the signal characteristics. Therefore, pre-processing (parameter optimization) and post-processing techniques should be viewed as complementary strategies rather than mutually exclusive alternatives.

\section{Differentiable Short-Time Fourier Transform}
\label{sec:3_dstft}
\subsection{Differentiable STFT with respect to window and hop lengths}

To define a differentiable formulation of the STFT with respect to its parameters -specifically, the window length and temporal positions- it is necessary to adapt the standard STFT definition given in Eq. \eqref{eq::eq1} to accommodate real-valued parameters. Note that differentiating the STFT with respect to the frame temporal position $t_\col$ is equivalent to differentiating with respect to the hop length $H_n = t_{n} - t_{n-1}, H_0 = t_0$. 

We begin by considering a base window function $\omega_L$ with compact support on $[-\support/2,\support/2]$. We then define a family of contracted windows as:
\begin{eqnarray}
\label{eq::tape_theta}
\tap (x,\winlength)  = \frac{\support}{\winlength} \tap_\support ( \frac{\support}{\winlength} x), \quad \forall 
\winlength \in ]0,\support].
\end{eqnarray}
By construction, $\tap(x, \winlength)$ is compactly supported on $[-\winlength/2, \winlength/2]$, such that
\begin{equation}
\label{eq::tape}
 \forall x \notin [-\winlength/2, \winlength/2] , \quad \tap(x,\winlength) = 0. 
\end{equation}
The normalization factor $\support/\winlength$ is crucial for preserving the $L^1$-norm  of the window as the parameter $\support$ is scaled to $\winlength$. This ensures a fair comparison of STFTs with varying window lengths. Specifically, if the input signal is a pure tone  $A \exp(j2 \pi a t)$, its STFT is $A \hat \omega (\eta-a,\winlength)$, where $\hat \omega$ is the Fourier transform. To ensure that the magnitude of the STFT at zero frequency shift is independent of $\winlength$, i.e., $\hat \omega(0, \winlength) = \int \omega(x, \winlength) dx$ is constant, the normalization in Eq. \eqref{eq::tape_theta} is necessary.

An example of such a window family derived from the normalized Hann function 
is:
\begin{equation}
\tap (x,\winlength) = \frac{1}{2\winlength} \left( 1+\cos \left( \frac{2 \pi x}{\winlength}\right) \right) 1_{ \lvert x \rvert \le \winlength/2}.
\end{equation}
Another common example is derived from the Gaussian function with variance $\sigma^2$ where $\sigma = \winlength / 6$:
\begin{equation}
\label{def_window}
\begin{aligned}   
\omega(x,\theta) = \frac{1}{\sigma} \exp \left(-\pi \left ( 
\frac{x}{\sigma}\right )^2 \right) 1_{ \lvert x \rvert \le \winlength/2}.
\end{aligned}
\end{equation}
The Gaussian window is widely used as its Fourier transform is also a Gaussian, $\exp (-\pi \sigma^2 \eta^2)$, preserving the amplitude of pure harmonics in the frequency domain. In this case, if we consider an effective support $ \support > \winlength = 6 \sigma $, then $\omega(x,\winlength) = \frac{L}{\winlength} \omega_L(\frac{L}{\winlength} x)$, with $\omega_L(x) = \exp (-\pi x^2)$ being very small at $x=\pm \support/2$. While the Gaussian window is not strictly compactly supported, its effective support allows for practical implementation. The differentiability at the exact boundaries of the compact support is a theoretical point that we will not delve into further in this work.

We can now define a generalized form of the STFT, taking into account variable window length and temporal position:
\begin{equation}
\label{eq::02}
    \begin{aligned}
\spec_{\tap}(t,m,\winlength) = \sum_{k\in\dZ} 
\tap( k-t,\winlength) s[ k ]  e^{- \frac{j2\pi km}{\support}}.
    \end{aligned}
\end{equation}
Then, considering $N$ times indices and $M = L/2+1$ frequency indices, one introduces the operator $\spec$ as:
\begin{eqnarray}
\label{eq::dstft}
\begin{aligned}
\spec & : \mathbb{R}^N\times ]0,L]^{M\times N} \mapsto \mathbb{C}^{M\times N}\\
         & \Omega = (t_n, \winlength_{m,n})_{m,n}
         \rightarrow (\spec_{\tap}(t_n,m,\winlength_{m,n}))_{m,n},
\end{aligned}
\end{eqnarray}
where $\Omega = (t_n)_{n=0}^{N-1} \cup (\winlength_{m,n})_{m=0, n=0}^{M-1, N-1}$ represents the set of trainable parameters. In this formulation, the window length $\winlength_{m,n}$ can vary with both time and frequency indices, while the temporal positions $t_\col$ are real-valued
and depend on the time index only. All these parameters are real-valued. The operator $\spec$ yields an $M\times N$ matrix, with rows corresponding to frequency and columns to time. 
If the bidimensional window function $(x,\winlength) \mapsto \tap (x,\winlength)$ is differentiable for all  $(x,\winlength)$,  then the modified STFT is differentiable with respect to both window and hop lengths (or temporal positions). In the following, we assume this property holds and refer to this modified STFT as DSTFT (for differentiable STFT).

The maximum frequency resolution in Eq. \eqref{eq::02} is determined by the support $\support$ (the size of the DFT), as the effective window of length $\winlength_{m,n} \le \support$ is implicitly zero-padded to this length. The parameter $\winlength_{m,n}$ governs the time resolution, analogous to $L$ in the classical STFT, by defining the temporal extent of the signal segment analyzed for a local spectrum. The zero-padding operation does not alter the intrinsic frequency resolution, which is determined by the total length $\support$; rather, it provides an interpolation of the frequency spectrum whose resolution is controlled by the window length $\winlength_{m,n}$. Therefore, $\support$ should be primarily considered as an upper bound on the temporal resolution $\winlength_{m,n}$ that allows maintaining a consistent size for the frequency dimension of the spectrogram, which is essential for differentiability, as will be discussed later.

In our DSTFT definition, the window length can a priori be different for each time and frequency index, and the hop length varies with the time index only. However, it is also possible to consider a window length that varies only with frequency, $\winlength_m$ (as in the S-transform), or only with time, $\winlength_n$ (as in some versions of ASTFT \cite{Zhong}), or a constant window length, $\winlength$, as in the classical STFT. The same applies to the temporal positions. Note that the classical STFT defined in Eq. \eqref{eq::eq1} can be obtained by setting $\winlength_{\f,\col}=\support$ and $t_\col = t_0 + nH$, where $H \in \dN$ is the hop length. While we could define temporal positions that depend on both time and frequency indices, such a representation would be challenging to interpret as the signal samples involved in the DSTFT computation for a given time index would vary with the frequency index. Nevertheless, this could potentially be used as input for a learning algorithm, but this is beyond the scope of the present paper.

\subsection{Fixed-Overlap DSTFT}
In the preceding section, we assumed that the temporal positions of the tapering window, $t_\col$, are independent of the window lengths, resulting in a constant number of time frames (columns) in the STFT representation. This configuration is particularly useful when the STFT serves as input to algorithms that require a fixed-size representation, such as neural networks. However, in certain applications, this assumption may not always be appropriate.
For instance, in spectrogram visualization, a common practice is to define a fixed overlap ratio $\alpha$ (e.g., 50\%) and make the temporal positions of the window dependent on the corresponding window lengths through the relationship $ t_\col = \col \alpha \winlength_{\col} $. It is important to note that with this setting, the number of time frames increases as $\winlength_{\col}$ decreases:

\begin{equation}
\label{eq::03}
\begin{aligned} 
\spec_{\tap}(\col \alpha \winlength_n,\f,\winlength_{\col}) = \sum_{k\in\dZ} 
\tap( k-\col \alpha \winlength_n,\winlength_n) s[ k ]  e^{- \frac{j2\pi k \f}{\support}}.
\end{aligned}
\end{equation}

\section{Analytical Expressions for partial derivatives}
\label{sec:4_grad}
In the preceding section, we introduced the DSTFT and established its differentiability with respect to its parameters, assuming a well-designed window function. This implies that the partial derivatives $\frac{\partial \spec}{\partial\winlength_{\f,\col}}$  and  $\frac{\partial \spec}{\partial t_\col}$ are well-defined and finite. In this section, we derive the analytical expressions for these derivatives and discuss how these formulas can be integrated into more general applications.

\subsection{Partial Derivatives with Respect to Window Length}
\label{sec:partial}
For notational simplicity, we denote the partial derivative of a function $f$ with respect to a variable $x$ as $\partial_x f$. 
Considering the case where the window length $ \winlength_{\f,\col}$  varies with both frequency index $\f$ and time index $\col$, the partial derivative of the DSTFT output $\spec(\Omega)$ with respect to $\winlength_{m,n}$ is given by:
\begin{equation}
\label{eq::eq04_v1}
\begin{aligned}
\partial_{\winlength_{\f,\col}} \spec (\Omega)_{\f', \col'}
&=\sum_{k\in\dZ} 
\partial_{\winlength_{\f,\col}} \tap (k-t_{\col'},\winlength_{\f',\col'}) s\left[ k \right]  e^{- \frac{j2\pi k \f'}{L}}\\
&= \spec_{\partial_\winlength \tap} (t_{\col'},m',\winlength_{\f',n'}) \delta_{\f,\f'} \delta_{n,n'}, 
\end{aligned}
\end{equation}
hence we denote:
\begin{equation}
\label{eq::eq04}
\begin{aligned}
\partial_{\winlength_{\f,\col}} \spec (\Omega)
&= (\spec_{\partial_\winlength \tap} (t_{\col'},\f',\winlength_{\f',n'}) \delta_{\f,\f'} \delta_{n,n'} )_{\f',n'}, 
\end{aligned}
\end{equation}
where $\delta_{\f,\f'}$ is the Kronecker delta, which equals $1$ if $\f=\f'$ and $0$ otherwise. Eq. \eqref{eq::eq04} represents a matrix where only the coefficient at time index $n$ and frequency index $m$ is non-zero. This non-zero coefficient corresponds to the STFT of the signal $s[k]$ computed using the partial derivative of the tapering function with respect to window length, $\partial_\winlength \tap$, instead of $\tap$ itself, evaluated at the specific time $t_\col$ frequency $m$ with window length $\winlength_{m,n}$.

In the case of frequency-only varying window length, $\cal S$ in defined
on $\dR^N\times ]0,L]^N$ and the partial derivative of the DSTFT 
with respect to window length $\winlength_{\col}$ is the following 
matrix:
\begin{equation}
\label{eq::eq05}
 \partial_{\winlength_{\col}} \spec (\Omega) = ( \spec_{\partial_{\winlength}\tap} (t_{\col'},m,\winlength_{n'}) \delta_{n,n'})_{m,n'}.
\end{equation}
In the case of a time-only varying window length, $\winlength_{n}$, where $\spec$ is defined on $\dR^N \times ]0,L]^M$, the partial derivative of the DSTFT with respect to $\winlength_{n}$ is the following matrix:
\begin{equation}
\label{eq::eq06}
 \partial_{\winlength_{\f}} \spec (\Omega) = (  \spec_{\partial_{\winlength}\tap} (t_{\col},m',\winlength_m) \delta_{m,m'})_{m',n}
\end{equation}
Finally, for a constant window length $\winlength$, where $\spec$ is defined on $\dR^N \times ]0,L]$, we obtain:
\begin{equation}
\label{eq::eq07}
\partial_\winlength \spec (\Omega) =   
( \spec_{\partial_{\winlength}\tap} (t_{\col},m,\winlength))_{m,n}.
\end{equation}

\subsection{Partial Derivatives with Respect to Window Temporal Position and Hop-Length}
\label{sec:partial2}
Following a similar approach to the previous section, and noting that the temporal position of the $n^{th}$ frame, $t_\col$, depends solely on the time index $\col$, we directly obtain the partial derivative of the DSTFT output with respect to $t_\col$ as:
\begin{equation}
\label{eq::eq08}
 \partial_{t_{\col}} \spec (\Omega) 
 = - (\spec_{\partial_x \tap} (t_{\col'},m,\winlength_{m,n'}) \delta_{n,n'})_{m,n'}.
\end{equation}
This represents a matrix where only the coefficient at time index n is non-zero, corresponding to the STFT of the signal using the partial derivative of the tapering function with respect to time, $\partial_x \tap$, evaluated at time $t_\col$, frequency $m$, and window length $\winlength_{m,n}$.
 
In the case of time-varying hop length $H_n$, where $H_n = t_{n} - t_{n-1}$ and $H_0 = t_0$, we remark that $t_n = \sum_{i=0}^{n} H_i$. We can use the chain rule to rewrite the partial derivative with respect to $t_n$:
\begin{equation}
\label{eq::eq08bis}
\begin{aligned}
 \partial_{H_{\col}} \spec (\Omega) &= 
 (\partial_{H_{\col}} \spec_{\tap}(t_{n'},m',\winlength_{m',n'}))_{m',n'}\\
 &= (-\spec_{\partial_x \tap}(t_{n'},m',\winlength_{m',n'}) 1_{x > 0} (n'- n))_{m',n'}\\
\end{aligned}
\end{equation}
where  $1_{x > 0}$ is the indicator function of $\mathbb{R}_*^+$.

For a constant hop length $H$, where $ t_\col = t_0 + \col H$, the partial derivative with respect to $H$ can be derived as:
\begin{equation}
\partial_H \spec (\Omega) =  ( -n \spec_{\partial_x \tap} (t_{\col},m,\winlength_{m,n}) )_{m,n}.
\end{equation}

\begin{remark}
An interesting characteristic of our DSTFT formulation is that all the derived partial derivative expressions retain the same structural form as the forward DSTFT. This implies that each derivative can be computed by performing a DSTFT operation using a modified tapering function (either $\partial_\winlength \tap$ or $\partial_x \tap$).
\end{remark}


\subsection{Backpropagation Formulas}
\label{sec:back_prop}
The partial derivative expressions derived in the previous subsections enable the computation of the gradient of any almost everywhere smooth differentiable scalar loss function with respect to the DSTFT parameters. To minimize such a loss function, gradient descent techniques can be employed in conjunction with the backpropagation algorithm \cite{werbos1990backpropagation}. Backpropagation provides an efficient method for calculating the gradient of a scalar loss function with respect to the parameters of a given function by recursively applying the chain rule. While widely utilized in deep learning \cite{lecun2015deep} for training neural networks, its applicability extends to computing partial derivatives of various functions.

Let $\loss$ be an almost everywhere smooth differentiable scalar loss function that depends on the STFT output $\spec$, which is a function of the parameter set $\Omega$. We can express this as $\loss \circ \spec (\Omega)$,  where $\circ$ denotes the composition of functions.
To optimize $\loss$ with respect to the DSTFT parameters, we can use gradient descent. The following derives the analytical expressions for backpropagation through the STFT.

First, we note that if we consider $\loss$ as a function of $\spec(\Omega)$, a matrix of $\mathbb{C}^{M \times N}$, then $\loss$ is a function from $\mathbb{C}^{M\times N}$ in $\dR$, and its derivative with respect to $\spec(\Omega)$, denoted by $\partial_\spec \mathcal{L}(\spec(\Omega))$, is a linear operator from $\mathbb{C}^{M\times N}$ to $\dR$, defined as:
\begin{equation}
\partial_\spec \loss (\spec (\Omega)) (C) =  \sum_{m,n} \frac{\partial \loss (\spec (\Omega))}
{\partial (\spec (\Omega))_{m,n}} C_{m,n} \in \dR, \ \forall C \in \mathbb{C}^{M\times N}.
\end{equation}

In the general case of a window length $\winlength_{m,n}$ varying with both frequency and time, we obtain the gradient with respect to the window length as:
\begin{eqnarray}
\label{eq::eq10}
 \begin{aligned}
 \partial_{\winlength_{\f,\col}} (\loss \circ \spec) (\Omega) &= 
 \partial_\spec \loss (\spec (\Omega))  \partial_{\winlength_{\f,\col}} \spec (\Omega)  \\
 &= \sum_{k,p} \frac{\partial \loss (\spec (\Omega))} {\partial (\spec (\Omega))_{k,p}} (\partial_{\winlength_{\f, \col}} \spec (\Omega))_{k,p}\\
 &=  \frac{\partial \loss (\spec (\Omega))} {\partial (\spec (\Omega))_{\f,\col}} \spec_{\partial_\winlength \tap} (t_{\col},\f,\winlength_{\f,\col}).
 \end{aligned}
 \end{eqnarray}
 Similarly, the gradient with respect to the temporal position $t_\col$ is:
\begin{eqnarray}
\label{eq::eq11}
 \begin{aligned}
\partial_{t_{\col}} (\loss \circ \spec) (\Omega) &= 
 \partial_\spec \loss (\spec (\Omega))  \partial_{t_\col} \spec (\Omega)  \\
 &= \sum_{k,p} \frac{\partial \loss (\spec (\Omega))} {\partial (\spec (\Omega))_{k,p}} (\partial_{t_\col} \spec (\Omega))_{k,p}\\
 &= - \sum_{k} \frac{\partial \loss (\spec (\Omega))} {\partial (\spec (\Omega))_{k,\col}} \spec_{\partial_x \tap} (t_{\col},k,\winlength_{k,\col}),
 \end{aligned}
 \end{eqnarray}
For the time-varying hop length $H_\col$, we have the relationship derived earlier:
\begin{eqnarray}
\label{eq::eq12}
 \begin{aligned}
 \partial_{H_{\col}} (\loss \circ \spec) (\Omega) 
 &= 
 \partial_\spec \loss (\spec (\Omega))  \partial_{H_\col} \spec (\Omega).
 \end{aligned}
 \end{eqnarray}

\noindent When the window length is time-only varying, $\winlength_{n}$, the gradient is:
\begin{eqnarray}
\label{eq::eq13}
 \begin{aligned}
 \partial_{\winlength_{\col}} (\loss \circ \spec) (\Omega) &= 
 \partial_\spec \loss (\spec (\Omega))  \partial_{\winlength_{\col}} \spec (\Omega)  \\
 &= \sum_{k} \frac{\partial \loss (\spec (\Omega))} {\partial (\spec (\Omega))_{k,n}} 
\spec_{\partial_{\winlength}\tap} (t_{\col},k,\winlength_{\col}).
\end{aligned}
\end{eqnarray}
When the window length is only frequency varying, $\winlength_{m}$, the gradient is: 
\begin{eqnarray}
\label{eq::eq14}
\begin{aligned}
 \partial_{\winlength_{\f}} (\loss \circ \spec) (\Omega) &= 
 \partial_\spec \loss (\spec (\Omega)) \partial_{\winlength_{\f}} \spec (\Omega)  \\
 &= \sum_{p} \frac{\partial \loss (\spec (\Omega))} {\partial (\spec (\Omega))_{m,p}} \spec_{\partial_{\winlength}\tap} (t_p,m,\winlength_m).
\end{aligned}
\end{eqnarray}
In the classical STFT case with constant window length $\winlength$ and constant hop length $H$,  the gradients are:
\begin{eqnarray}
\label{eq::eq15}
\begin{aligned}
 \partial_{\winlength} (\loss \circ \spec) (\Omega) &= 
\partial_\spec \loss (\spec (\Omega)) \partial_{\winlength} \spec (\Omega)  \\
 &=\sum_{k,p} \frac{\partial \loss (\spec (\Omega))} {\partial (\spec (\Omega))_{k,p}} \spec_{\partial_{\winlength}\tap} (t_p,k,\winlength),
 \end{aligned}
 \end{eqnarray}
 and
 \begin{eqnarray}
 \begin{aligned}
 \partial_{H} (\loss \circ \spec) (\Omega) &= 
 \partial_\spec \loss (\spec (\Omega)) \partial_{H} \spec (\Omega)  \\
 &=-\sum_{k,p} \frac{\partial \loss (\spec (\Omega))} {\partial (\spec (\Omega))_{k,p}} p\spec_{\partial_x \tap} (t_p,k,\winlength_{k,p}).
\end{aligned}
\end{eqnarray}

\begin{remark}
These analytical expressions provide considerable flexibility, allowing for the differentiation of any scalar function $\loss$ of the DSTFT outputs with respect to its tuning parameters. Notably, all backpropagation computations can be efficiently implemented using matrix multiplications, with the involved matrices having the same dimensions as those in the forward propagation. This leads to faster gradient calculations with exact values, offering an advantage over traditional automatic differentiation tools.
\end{remark}
\begin{remark}
These backpropagation formulas enable the use of gradient descent optimization for minimizing any differentiable cost function. Unlike previous methods such as adaptive STFT, which typically evaluate the cost function over a predefined discrete set of window sizes to select the optimal one, our approach employs gradient-based optimization algorithms like stochastic gradient descent to directly minimize the loss. It is important to acknowledge that the inherent challenges of convergence and convexity associated with gradient-based methods in machine learning remain applicable.
\end{remark}

\section{Computational Aspects}
\label{sec:computation}
\subsection{Numerical Implementation}

To implement the DSTFT numerically, we address the infinite summation by leveraging the finite support of the tapering function $\tap(x, \winlength)$, which is non-zero only for $x \in [-\support/2, \support/2]$. This allows us to restrict the summation over $k \in \dZ$ to this finite interval. To handle non-integer temporal positions $t_\col$, we decompose $t_\col$ into its integer part $\lfloor t_\col \rfloor$ and its fractional part $\{t_\col\} = t_\col - \lfloor t_\col \rfloor$.
This leads to the following practical expression for the DSTFT:
\begin{equation}
\begin{aligned}
&\spec_{\tap}(t_\col,\f,\winlength_{\f, \col}) = 
\sum_{k\in\dZ} 
\tap(k - t_\col, \winlength_{\f,\col}) s[ k ]  e^{- \frac{2j\pi k \f}{\support}} \\
&= \sum_{k=-\support/2+1}^{\support/2} \tap(k - \{t_\col\}, \winlength_{\f,\col}) s\left[ \lfloor t_\col \rfloor +k \right]  e^{- \frac{2j\pi}{\support}(k+\lfloor t_\col \rfloor)\f} \\
&= e^{- \frac{2j\pi (\lfloor t_\col \rfloor )\f}{\support}}  \\
&\sum_{k=-\support/2+1}^{\support/2} \tap (k - \{t_\col\}, \winlength_{\f,\col}) s\left[ \lfloor t_\col \rfloor + k \right]  e^{- \frac{2j\pi k \f}{\support}}.
 \end{aligned}
\end{equation}

This formulation effectively considers the signal samples centered around the integer time index $\lfloor t_\col \rfloor$ and accounts for the fractional offset $\{t_\col\}$ through a shift in the argument of $\tap$ and a phase factor $e^{-j2\pi (\lfloor t_\col \rfloor ) \f / \support}$ in the exponential term. The summation limits are set from $-\support/2+1$ to $\support/2$ because $0 \leq \{t_\col\} < 1$, and the tapering function $\tap$ is zero at the boundaries of its support. Consequently, the analysis window function remains  a continuous and differentiable function of real-valued window lengths and temporal frame positions, and we evaluate this analysis window on discrete time indices.

Within the differentiability framework established in the preceding section, we assumed a fixed size for the spectrogram, resulting in a constant number $N$ of time frames (columns). However, for a fixed-overlap implementation of the DSTFT where the number of time frames adapts to the window length, the number of columns with temporal overlap with the signal can vary. To accommodate this variability while adhering to the previously defined differentiability framework, we employ zero-padding of the input signal. This ensures that all windows potentially encompassing at least one original sample are considered in the analysis.
This approach allows for an STFT with a number of time frames that adjusts based on the chosen time resolution, analogous to the classical STFT when maintaining a constant overlap ratio as the window length changes. 
Specifically, a differentiable fixed-overlap DSTFT with a time-varying window length can be defined by setting the initial time frame $t_0=0$ and subsequent time frame positions as $t_\col = t_{\col-1} + \alpha \winlength_{\col-1}$ for $\col > 0$, where $\alpha$ represents the constant overlap factor.
 
The codes and experimental results are available on our GitHub repository: \url{https://github.com/maxime-leiber/dstft}.

\subsection{Computational Complexity}
\label{sec:comput_comp}
For any set of parameters $\Omega$,  the DSTFT output $\spec(\Omega)$ can be computed in $O(N \support^2)$ operations. However, when the window length depends only on the time index ($\winlength_{m,n} = \winlength_n$) or is constant ($\winlength_{m,n} = \winlength$), the Fast Fourier Transform (FFT) can be utilized at each time index, reducing the computational cost to $O(N \support \log \support)$, similar to the classical STFT. This efficient computational complexity makes the time-varying window length approach particularly advantageous for real-time applications requiring fast processing.

Regarding the gradient computation, as detailed in Sec. \ref{sec:back_prop}, the backward pass has a computational complexity equivalent to the forward pass: $O(N\support \log \support)$ when using the FFT, and $O(N\support^2)$ otherwise.
This is because all expressions in Sec. \ref{sec:back_prop} involve the term $\partial_\spec \mathcal{L}(\spec(\Omega))$, which is computed once, and DSTFTs based on differentiated window functions, which have the same complexity as the original DSTFT.

To provide a direct comparison with classical discrete optimization techniques, let $A$ denote the cardinality of the hop-length set and $B$ that of the window lengths used in discrete optimization approach, and let $P$ be the number of iterations in the gradient descent technique.  Table \ref{tab:table2} summarizes the approximate computational complexity of these two types of methods for various scenarios.
\begin{table}[h]
\centering
\caption{Comparison with discrete optimization approach}
\label{tab:table2}
\begin{tabular}{c|c|c}

 Case &  Discrete approaches &  DSTFT-based   \\ 
 \hline 
 \\
  Constant                   &   $O(A B N  
 \support \log \support )$ &  $O(2  P  N \support \log \support)$  \\
  Time-varying                 &    $O((AB)^N  N  \support \log \support )$ &  $O(2  P  N  \support \log \support)$  \\
  TF-varying  &   $O(A^N B^{NM} N  \support^2)$   &  $O(2 P  N  \support^2)$ \\
 
\end{tabular}
\end{table}

As shown in Table \ref{tab:table2}, the computational complexity of DSTFT-based approach scales with the number of gradient iterations $P$. In contrast, the discrete optimization method requires evaluating the cost function for every combination of candidate parameters within the defined search space. This leads to an exponential increase in complexity for time-varying and, particularly, TF varying parameters, often rendering such exhaustive searches computationally prohibitive or even infeasible.

This comparison highlights a key advantage of our DSTFT framework. For the case of constant window and hop lengths, the computational complexity is comparable to that of discrete search methods. More significantly, for the more complex and practical scenarios of time-varying and TF varying parameters, our DSTFT-based approach offers a computationally tractable alternative to discrete search algorithms.

\section{Applications: Representation-Driven Optimization}
\label{sec:5_exp}
In the following section, we will explore applications of DSTFT-based approach, 
in three different optimization problems, aiming at enhancing the TF representation in different contexts. 
The first two examples are on simulated signals 
and involve  respectively TF varying window length and time varying hop and window 
lengths. Though these examples were already studied in \cite{dastft} 
and in \cite{dstft2}, more details are provided here on relevant algorithmic 
choice and on implementation. Furthermore, the novel formalism detailed in previous sections provides a clearer view than the algorithmic approaches  
proposed in \cite{dastft} and \cite{dstft2}. Finally, an novel illustration 
of the proposed framework on vibration signals conclude the section. 

\subsection{Time and Frequency Varying Window Length} 
\label{sec:TF_varying}
We consider a simulated signal composed of three components: a non-stationary component, a stationary component that is close in frequency to the first component for a certain duration, and a transient component. Our focus in this example is on optimizing the window length while maintaining a constant hop length. All simulations in this study employ Hann windows.

\begin{figure}
    \centering
    \begin{tikzpicture}
        \node[inner sep=0pt] (img1) {\includegraphics[width=4.5cm]{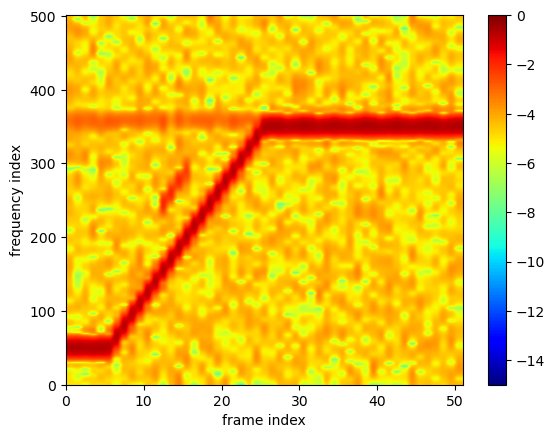}};
        \node[inner sep=0pt] (img2) at (img1.east) [right=0cm] {\includegraphics[width=4.5cm]{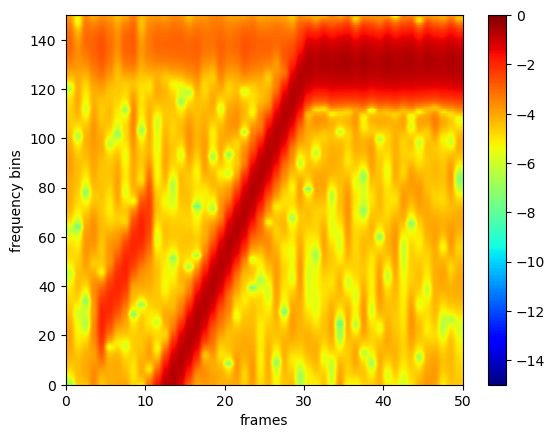}};
        \node[inner sep=0pt] (img3) at (img1.south) [below=0.0cm] {\includegraphics[width=4.5cm]{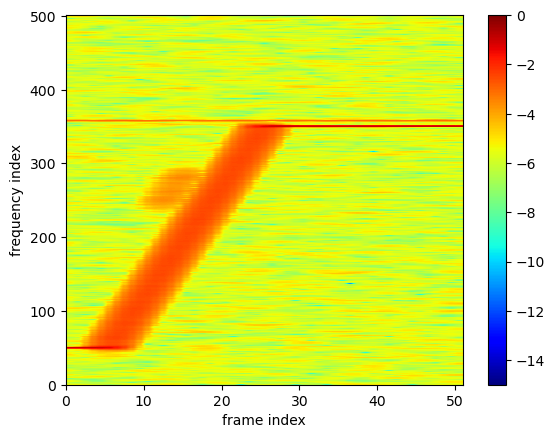}};
        \node[inner sep=0pt] (img4) at (img3.east) [right=0cm] {\includegraphics[width=4.5cm]{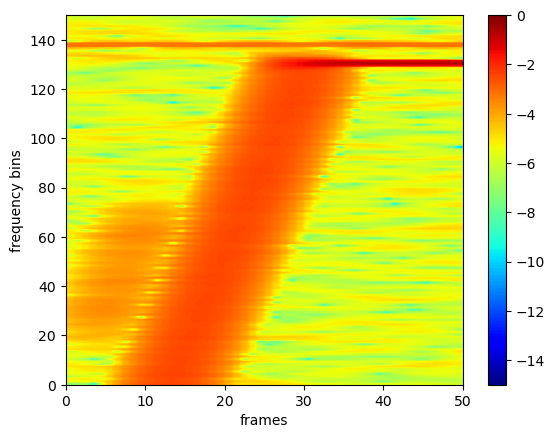}};
        \node[inner sep=0pt] (img5) at (img3.south) [below=0.0cm] {\includegraphics[width=4.5cm]{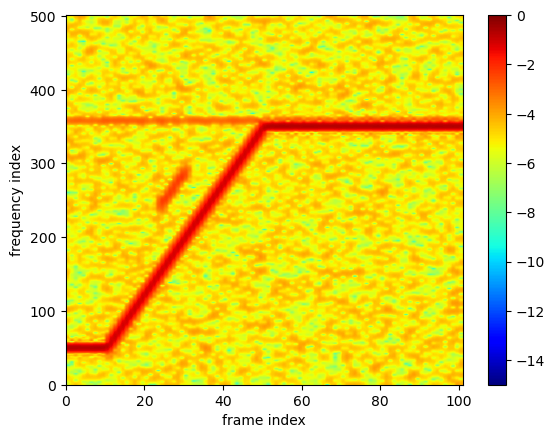}};
        \node[inner sep=0pt] (img6) at (img5.east) [right=0cm] {\includegraphics[width=4.5cm]{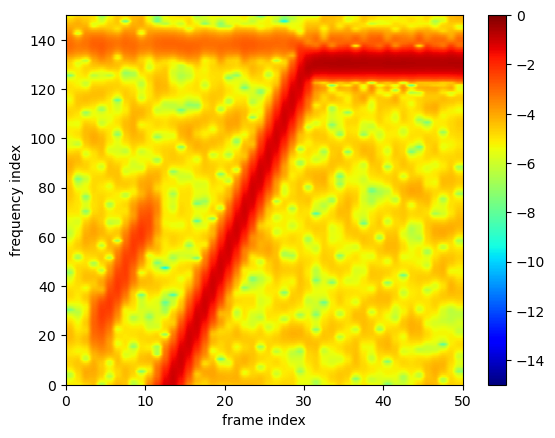}};
    \end{tikzpicture}
    \caption{Spectrograms (left) and zoomed-in spectrograms (right) with  respectively from top to bottom small window of length 100, long window of length 1000, and constant-window DSTFT.}
    \label{fig:fig3}
\end{figure}

First, we illustrate the limitations of using a single, fixed window length. When the STFT is computed with a short window, it provides good temporal localization but poor frequency resolution, particularly for the transient event. Additionally, the non-stationary and stationary components are not well distinguished (see the first row of Fig. \ref{fig:fig3}). Conversely, a long window allows for better separation of the non-stationary and stationary components but obscures the transient event (see the second row of Fig. \ref{fig:fig3}).

\begin{figure}
    \centering
    \begin{tikzpicture}
        \node[inner sep=0pt] (img1) {\includegraphics[width=5cm]{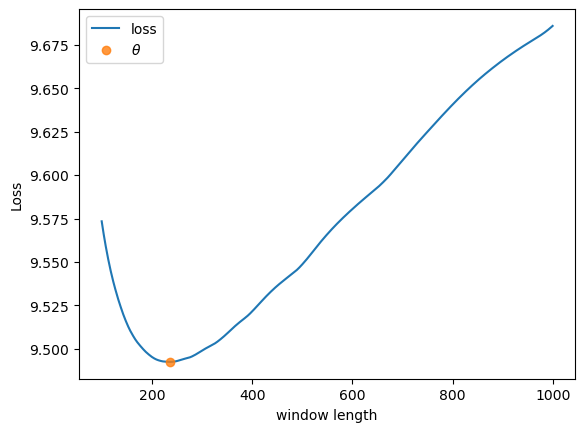}};       
    \end{tikzpicture}
    \caption{Loss function corresponding to Eq. \eqref{def:entropy} with respect to window length}
    \label{fig:fig3c}
\end{figure}

To address these limitations, we first propose optimizing the DSTFT with a single window length parameter, $\Omega = \theta$, by minimizing the Shannon entropy loss: 
 \begin{equation}
 \label{def:entropy}
 \ent \circ \spec (\theta) = - \sum_{\f, \col} p_{\f,\col} 
 \log (p_{\f, \col}),
 \end{equation}
 where $p_{\f,\col} = \frac{| \spec_{\tap}(t_\col,\f,\theta)|}
 {\sum_{k, l} | \spec_{\tap}(t_l,k,\theta)|}$. The Shannon entropy $\ent$ 
 is differentiable  with respect to $\spec (\theta)$ provided that
 $\spec_\tap(t_n,m,\theta)$ is non-zero for all indices $m$ and $n$. 
The Shannon entropy criterion is commonly used in the literature as it is related to minimizing interference in the TF plane \cite{baraniuk2002measuring}. Other entropy measures, such as R\'enyi entropy could also be considered \cite{meignen2020use}. 
With a fixed hop length $H$ and a signal of length $L_s$, the number of time frames is set to $N = 1 + \lfloor \frac{L_s}{H} \rfloor$.
The optimization process converges to a spectrogram (displayed in the third row of Fig. \ref{fig:fig3}) that is optimal according to the entropy criterion. In this specific case, numerical analysis (see Fig. \ref{fig:fig3c}) suggests that the loss function is convex with respect to $\winlength$, ensuring the uniqueness of the optimum. Future work will focus on rigorously proving the convexity of this loss function.

\begin{figure}
    \centering
    \begin{tikzpicture}
        \node[inner sep=0pt] (img1) {\includegraphics[width=4.5cm]{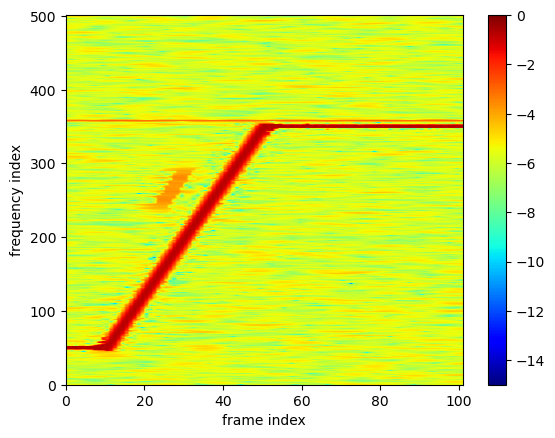}};
        \node[inner sep=0pt] (img2) at (img1.east) [right=0cm] {\includegraphics[width=4.5cm]{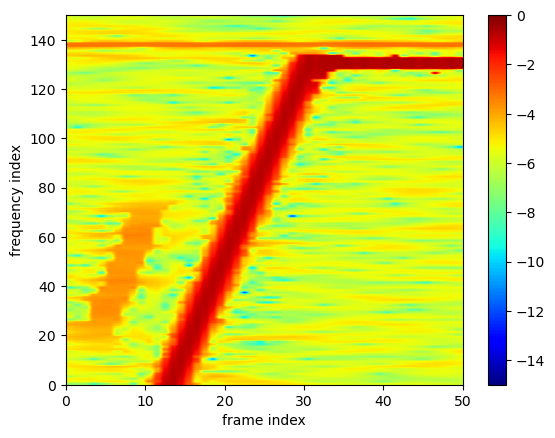}};
        \node[inner sep=0pt] (img3) at (img1.south) [below=0.0cm] {\includegraphics[width=4.5cm]{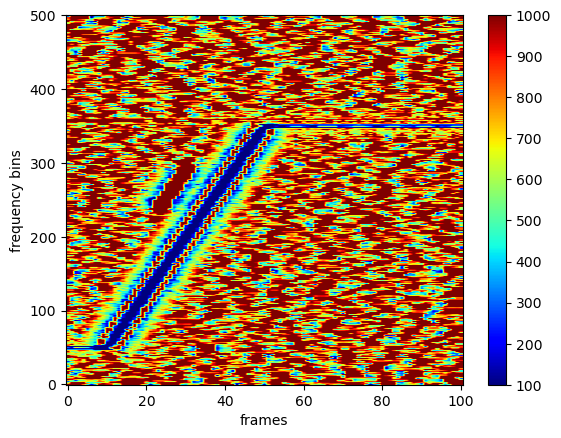}};
        \node[inner sep=0pt] (img4) at (img3.east) [right=0cm] {\includegraphics[width=4.5cm]{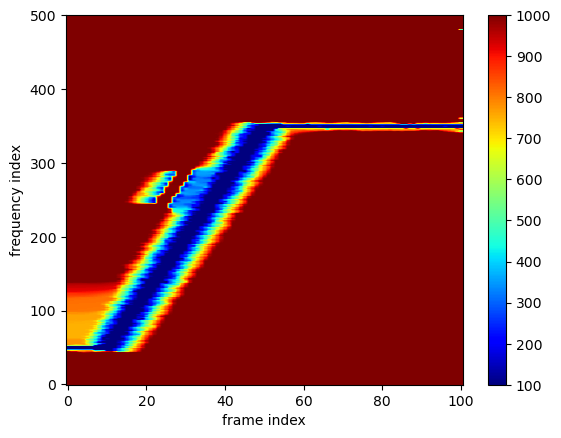}};        
    \end{tikzpicture}
    \caption{Top row: Spectrogram (left) and its zoomed-in section (right) computed with a TF varying window length. Bottom row: Visualization of the associated window length distribution, illustrating the effect of the regularization term (left: without, right: with).}
    \label{fig:fig3b}
\end{figure}

Regarding computational cost, a grid-search approach with a window length ranging from $100$ to $1000$ samples would require evaluating 901 discrete window lengths ($B=901$). In contrast, the DSTFT optimization typically converges within $P$ iterations, where $P$ depends on the choice of hyperparameters (e.g., learning rate, stopping condition) and the initial window length. Empirically, $P$ usually ranges from $10$ to $100$. Considering that each iteration involves a forward and backward pass, the computational cost is roughly $2P$ times that of a single STFT computation. Thus, in the worst case, the DSTFT optimization requires approximately $200$ forward/backward passes, which is significantly less than the $901$ evaluations required by grid search. Furthermore, the DSTFT optimization can find optimal window lengths that are not restricted to a discrete grid, potentially leading to more accurate results.

Next, we investigate the potential for further improvement in TFR by allowing the window length to vary with both time and frequency. For the considered simulated signal, finding a single constant window length that optimally represents all its diverse components in the spectrogram appears challenging. We therefore consider $\Omega = (\theta_{m,n})_{m,n}$ as the set of parameters to optimize the DSTFT. In this case, the Shannon entropy loss  $\ent \circ \spec (\Omega)$ is no longer necessarily convex. We thus consider minimizing the following criterion:
\begin{equation}
\label{eq:loss}
\tilde \loss(\Omega) = \ent \circ \spec (\Omega) + \lambda  \mathcal{R}(\Omega),
\end{equation}
where $\ent$ is the Shannon entropy, $\mathcal{R}(\Omega)$ is a regularization term that encourages neighboring windows in the TF plane to have similar lengths, thereby enhancing robustness to noise, and  $\lambda$ is an hyperparameter controlling the trade-off between these two terms. The rationale for using such a regularization term is that as $\lambda \rightarrow +\infty$, the optimal solution should approach a constant window length, for which the loss function has been observed to be convex (as illustrated in Fig. \ref{fig:fig3c}). In our simulations, we considered the following regularization term:
\begin{eqnarray}
\label{eq:def_R}
\mathcal{R}(\Omega) = \sum_{n,m} \sqrt{\sum_{(n',m') \in N_{n,m}} 
(\winlength_{n,m}-\winlength_{n',m'})^2},
\end{eqnarray}
where $N_{n,m}$ is a set of indices corresponding to the neighbors of the bin  $(n,m)$ in the TF plane. 
Note that $\mathcal{R}$ is differentiable with respect to $\Omega$, and the gradient of $\tilde \loss(\Omega)$ can be computed using the chain rule for the entropy term and direct differentiation for the regularization term. 
This regularization term is related to non-local total variation penalization, commonly used in image processing for its robustness to noise \cite{nltv}.

Results obtained with the DSTFT using a TF varying window length are shown in the first row of Fig. \ref{fig:fig3b} for the spectrograms. The corresponding window length values, without and with the regularization term ($\lambda = 10^{-3}$), are displayed in the second row of Fig. \ref{fig:fig3b}. This clearly demonstrates the necessity of using a regularization term to obtain meaningful optimal window lengths. This approach allows for precise localization of all signal components in both time and frequency, significantly enhancing the overall quality of the TFR by adapting the window length to both transient and stationary characteristics of the signal.

Regarding the computational cost for a TF varying window length, a grid-search approach involving window lengths ranging from $100$ to $1000$ samples would require testing $B=901^{101\times501}$ combinations, which is computationally intractable. In contrast, our DSTFT approach requires $2P$ iterations, where $P$ depends on hyperparameters such as the learning rate, stopping condition, the regularization parameter, and the initial window lengths. To facilitate convergence towards a global optimum, a practical strategy is to initialize the window lengths with the optimal constant window length obtained in the previous convex optimization step. Empirically, $P$ typically ranges from $50$ to $300$, resulting in a computational cost of at most $600$ times that of a single forward pass, which is negligible  compared to the cost of grid search.  
Furthermore, the proposed optimization technique is not restricted to a discrete set of window lengths, potentially leading to more accurate TFRs. To the best of our knowledge, this work presents a novel approach for computing TFRs associated with TF varying window lengths.

\subsection{Time-Varying Window and Hop-Lengths} 
\label{sec:hop_length}

\begin{figure}[t]
    \centering
    \begin{tikzpicture}
        \node[inner sep=0pt] (img1) {\includegraphics[width=4.5cm]{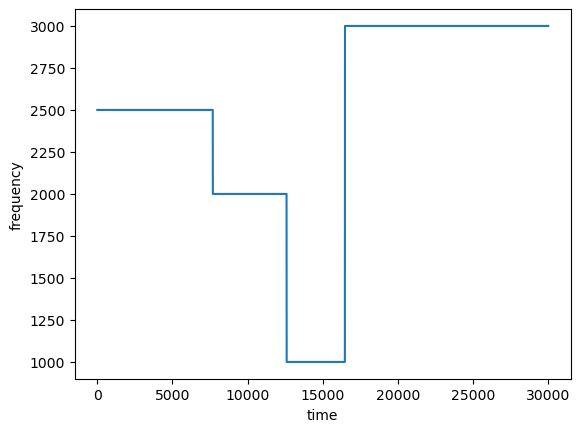}};
        \node[inner sep=0pt] (img2) at (img1.east) [right=0cm] {\includegraphics[width=4.5cm]{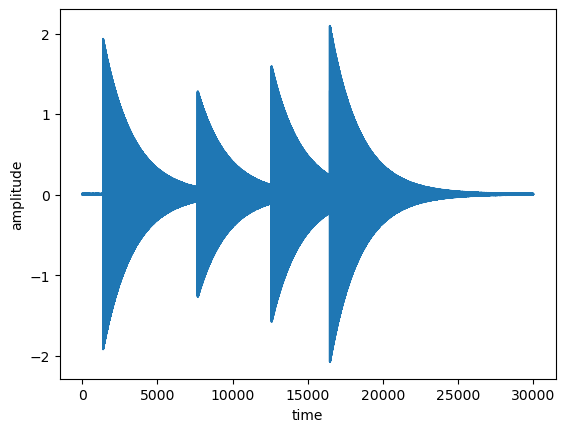}};
        \node[inner sep=0pt] (img3) at (img1.south) [below=0.0cm] {\includegraphics[width=4.5cm]{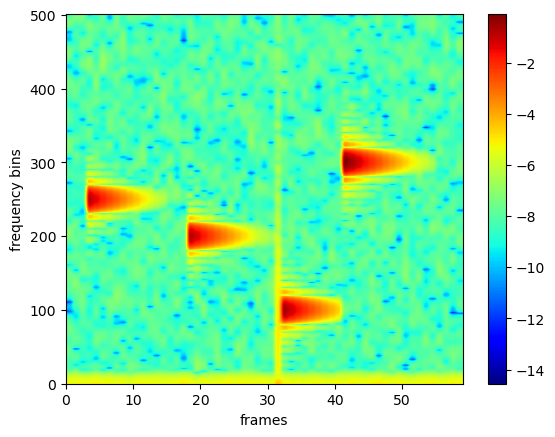}};
        \node[inner sep=0pt] (img4) at (img3.east) [right=0cm] {\includegraphics[width=4.5cm]{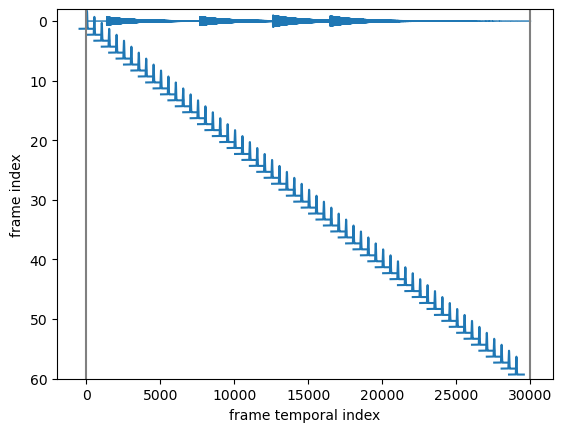}};
        \node[inner sep=0pt] (img5) at (img3.south) [below=0.0cm] {\includegraphics[width=4.5cm]{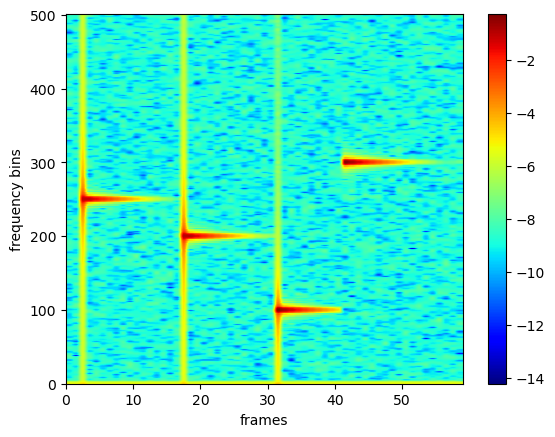}};
        \node[inner sep=0pt] (img6) at (img5.east) [right=0cm] {\includegraphics[width=4.5cm]{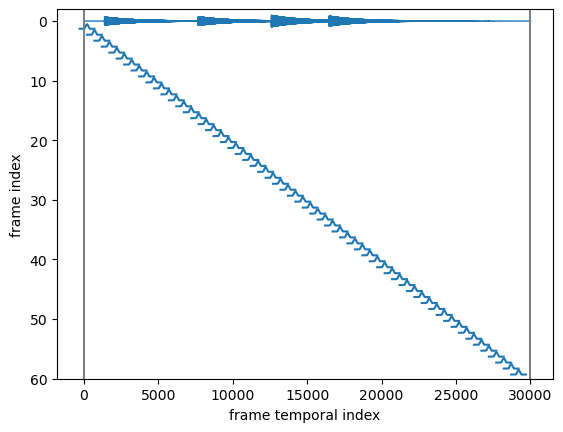}};
        \node[inner sep=0pt] (img7) at (img5.south) [below=0.0cm] {\includegraphics[width=4.5cm]{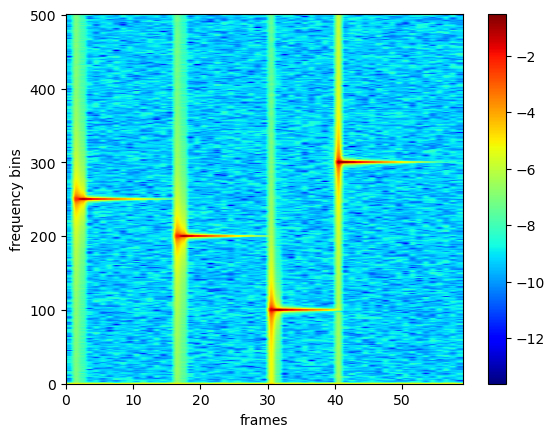}};
        \node[inner sep=0pt] (img8) at (img7.east) [right=0cm] {\includegraphics[width=4.5cm]{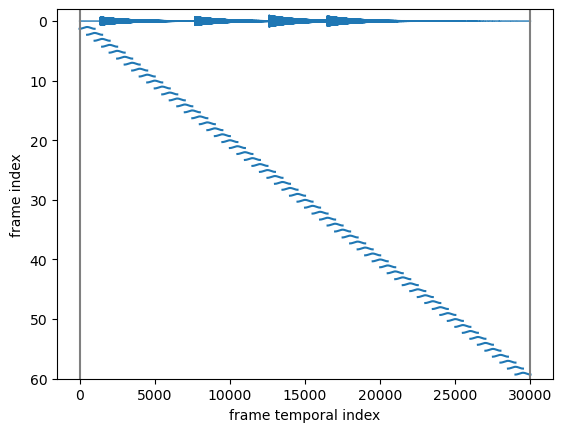}};
        \node[inner sep=0pt] (img9) at (img7.south) [below=0.0cm] {\includegraphics[width=4.5cm]{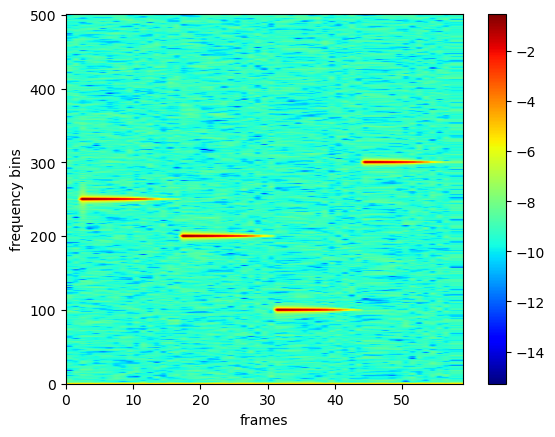}};
        \node[inner sep=0pt] (img10) at (img9.east) [right=0cm] {\includegraphics[width=4.5cm]{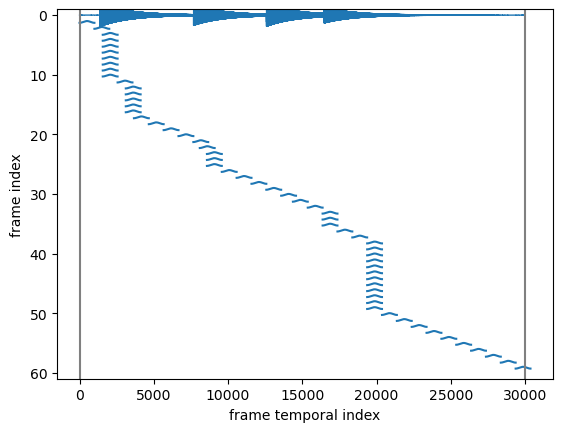}};
    \end{tikzpicture}
    \caption{In the first row, frequencies (left) and temporal signal (right). Spectrograms (left) and frame temporal position (right) with  respectively from second row to bottom small window of length 100, medium window of length 400, long window of length 1000, DSTFT with time-varying window and hop length.}
    \label{fig:fig4}
\end{figure}

This section investigates the optimization of the DSTFT when both window and hop lengths are varied, using a simulated signal. In this scenario, the set of parameters is $\Omega = (t_n, \theta_n)_n$. 
To demonstrate the benefits of optimizing the DSTFT with respect to these parameters, we focus on signals containing transient events (shocks) with varying frequencies and durations, as depicted in the first row of Fig. \ref{fig:fig4}. Gaussian white noise is added to the signal to achieve a signal-to-noise ratio (SNR) of 20 dB.

\begin{figure}[ht]
\centering
\begin{tikzpicture}[scale=1.7]
\draw[help lines, very thin, color=gray!50, dashed] (.1,0.3) grid (5.1,3.4);
\draw[domain=.4:5.1, smooth] plot (\x,{3.3+ .2*sin( 10 * \x r)}); 
\draw (.05,3.25) node[below, scale=1.5] {$s$};
\newcommand*{\rotatecurvearrowleft}{\mathbin{\rotatebox{90}{$\curvearrowleft$}}}
\draw (.45,2.85) node[scale=2] {$ \rotatecurvearrowleft $}; 
\draw (.05,2.8) node[scale=1.2] {$\times$};

\draw[domain=1.3:3.5, color=BleuBleu] plot (\x, { 2+  .5*(1-cos(2*pi*(\x-.4-.9)/2.2 r))  });
\draw[domain=0.:1.3, color=BleuBleu] plot (\x, 2);
\draw[domain=3.5:4., color=BleuBleu] plot (\x, 2);
\draw [BleuBleu, stealth-stealth](1.3, 1.85) -- (3.5, 1.85);
\draw (2.4,1.85) node[below, BleuBleu] {$\winlength_{\col}$};
\draw [BleuBleu](2.4, 2.) -- (2.4, 1.9);
\draw (2.4, 1.95) node[above, BleuBleu] {$t_\col$};
\draw (2.4, 2.15) node[above, color=BleuBleu ] { $\tap (x-t_n, \winlength_n )$ };


\draw[domain=2.8:4.8, color=darkgreen ] plot (\x, {1+  .5*(1-cos(2*pi*(\x-2.8)/2 r))  });
\draw[domain=1.:2.8, color=darkgreen] plot (\x, 1);
\draw[domain=4.8:5., color=darkgreen] plot (\x, 1);
\draw [darkgreen, stealth-stealth](2.8, .85) -- (4.8, .85);
\draw (3.8,.85) node[below, darkgreen] {$\winlength_{\col +1}$};
\draw [darkgreen](3.8, 1.) -- (3.8, .9);
\draw (3.8, .95) node[above, darkgreen] {$t_ {\col +1}$};
\draw (3.8, .2) node[above, color=darkgreen ] { $\tap (x-t_{n+1}, \winlength_{n+1} )$ };

\draw [dashed, darkred](2.4, 1.4) -- (2.4, 1.6);
\draw [dashed, darkred](3.8, 1.4) -- (3.8, 1.1);
\draw [stealth-stealth, color=darkred](2.4, 1.4) -- (3.8, 1.4);
\draw (3.1, 1.4) node[above, color=darkred] {  $H_{\col+1}$}; 

\end{tikzpicture}
\caption{The position of the tapering windows can smoothly shift along the time axis, while the window supports start at the integer part of the temporal position of the tapering windows.
}
\label{fig:fighop}
\end{figure}
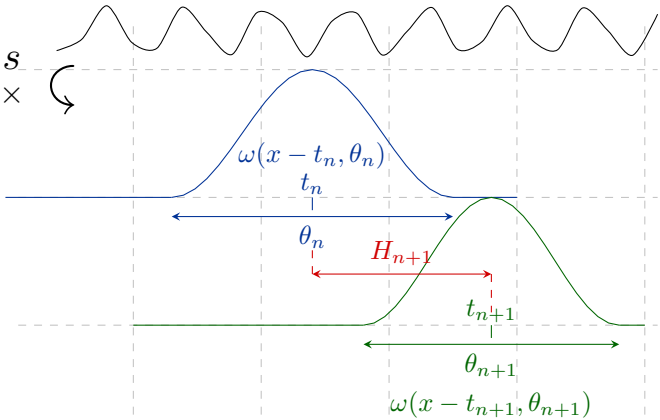

For such signals, employing frames uniformly localized along the time axis is suboptimal. Regardless of the chosen window length, energy leakage is observed in the spectrogram, as illustrated in the second, third, and fourth rows of Fig. \ref{fig:fig4}. We therefore propose optimizing the window positions and lengths in the DSTFT using gradient descent. However, this optimization requires careful consideration to ensure that the set of translated windows used to compute the spectrogram adequately covers the original signal. With this in mind, to promote energy concentration within each time frame, we aim to maximize the kurtosis of the frame spectrum \cite{Zhao} while also incorporating a regularization term to ensure adequate signal coverage. This leads to the following loss function:
\begin{equation}
\label{eq:loos_2}
\tilde \loss(\Omega) = \mathcal{K} \circ \spec (\Omega) + \lambda  
\mathcal{C}(\Omega),
\end{equation}
in which: 
\begin{equation}
\label{eq:kurtosis}   
\mathcal{K} \circ \spec (\Omega) = 
\frac{1}{N} \sum_{\col} \frac{\mathbb{E}_\f [| \spec (t_\col,\f,\theta_n)|^4]}
{\mathbb{E}_\f[| \spec (t_\col,\f,\theta_n)|^2]^2}.
\end{equation}
The number of time frames $N$ is initially determined by choosing an arbitrary fixed hop length $H$ and setting $ N = 1 + \lfloor \frac{L_s}{H} \rfloor$, where $L_s$ is the length of the signal.
The penalization term $\mathcal{C}(\Omega)$ is introduced to ensure that the set of adaptive windows provides sufficient coverage of the entire signal. A condition for adequate coverage is that the overlap between successive windows should satisfy:
\begin{equation}
\label{eq:1}
t_{n+1} - t_n \leq \frac{\theta_{n+1} + \theta_n}{2}.
\end{equation}
To encourage maximum spectrogram coverage, we consider the following penalization term:
\begin{equation}
\label{eq:coverage}
\begin{aligned}
\mathcal{C}(\Omega)=&\frac{1}{L_s}\sum_{\col=0}^{N-1} 
\min (t_{\col+1} -t_\col, \frac{\theta_{\col+1} +\theta_\col}{2})\\
&\hspace{1.2 cm} 1_{ x < L_s}(t_{n+1}-\frac{\theta_{n+1}}{2}) 
1_{ x > 0}(t_{n}+\frac{\theta_n}{2}).
\end{aligned}
\end{equation}
When $\mathcal{C}(\Omega)$ is significantly smaller than 1, it indicates that the set of translated windows does not fully cover the signal. Note that $\mathcal{C}(\Omega)$ is differentiable with respect to $\Omega$, 
anywhere in the parameter space.
To better understand the motivation for this penalization term, Fig. \ref{fig:fighop} illustrates the overlapping constraints between two successive windows.

As shown in the fifth row of Fig. \ref{fig:fig4}, appropriate time positioning of the frames leads to better energy concentration due to reduced spectral leakage. Observing the time axis of the resulting TFR, one can also notice that the distribution of the frames along the time axis is no longer uniform.

\begin{figure}[t]
    \centering
    \begin{tikzpicture}
        \node[inner sep=0pt] (img1) {\includegraphics[width=4.5cm]{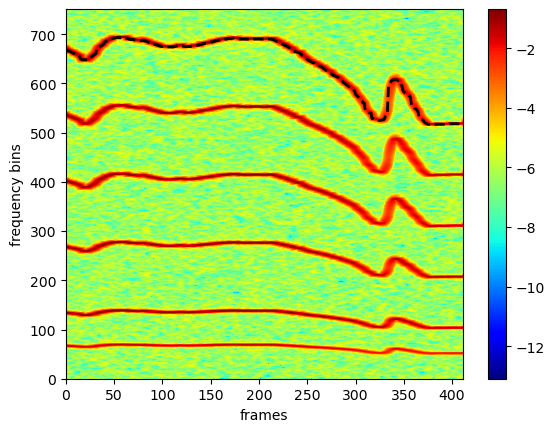}};
        \node[inner sep=0pt] (img2) at (img1.east) [right=0cm] {\includegraphics[width=4.5cm]{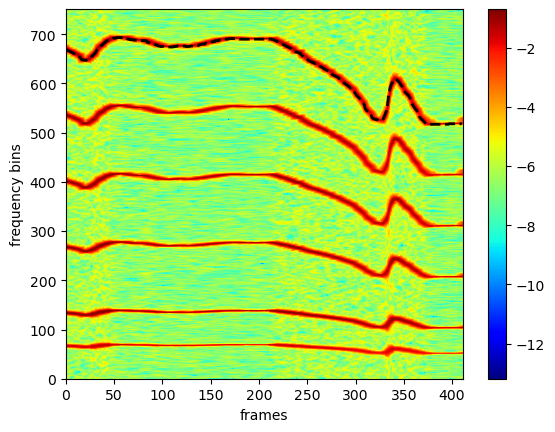}};
        \node[inner sep=0pt] (img3) at (img1.south) [below=0.0cm] {\includegraphics[width=4.5cm]{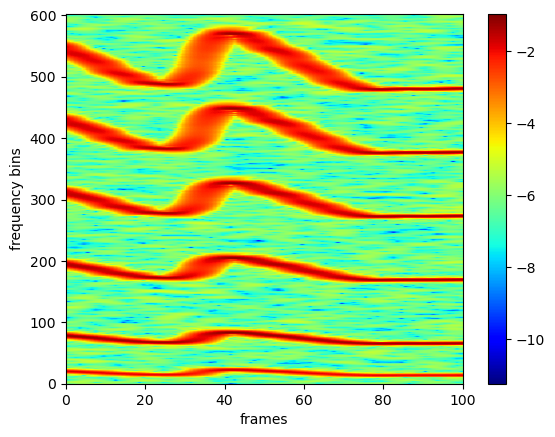}};
        \node[inner sep=0pt] (img4) at (img3.east) [right=0cm] {\includegraphics[width=4.5cm]{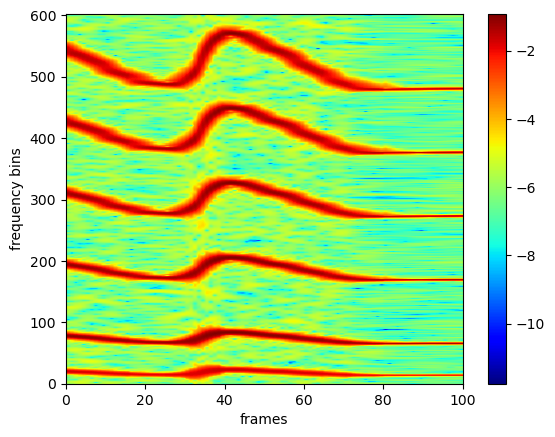}};
    \end{tikzpicture}
    \caption{Single window (left) and time-varying window (right) Spectrograms using DSTFT with respectively from top to bottom spectrogram with estimated instantaneous frequency, and zoomed-in spectrogram.}
    \label{fig:fig5}
\end{figure}

\subsection{Window Length Optimization for Frequency Tracking from a Vibration Signal} 
\label{sec:vibration}

In this section, we consider optimizing the window length in DSTFT when applied to a real-world multi-harmonic vibration signal obtained from an aircraft engine. The primary objective is to estimate the rotational speed of the shaft, which corresponds to the main harmonic frequency, using techniques such as ridge tracking in the TF plane, leveraging the signal's multi-harmonic nature as described in \cite{leclere2016multi}. A key challenge in this task is to locally determine an appropriate window length that can accurately track rapid frequency variations while maintaining sufficient frequency resolution. It is worth noting that the studied multi-harmonic signal is synthesized by considering multiples of the fundamental frequencies recorded during an actual aircraft flight, thus providing access to the ground truth of the main harmonic frequency. Additionally, white Gaussian noise is added to the signal to simulate realistic conditions.

In this example, our goal is to optimize the DSTFT, specifically focusing on time-varying window lengths, by minimizing the Shannon entropy loss. We then use the resulting TFR to estimate the main harmonic frequency. The first row of Fig. \ref{fig:fig5} displays the optimal spectrogram obtained using a single, constant window length (left) and the optimal spectrogram obtained using a time-varying window length (right), both achieved by minimizing the Shannon entropy. The second row of Fig. \ref{fig:fig5} shows zoomed-in views of these spectrograms in non-stationary regions, clearly illustrating the benefits of employing a time-varying window length during the optimization process for enhanced resolution.

Enhancing the TFR is crucial in numerous vibration spectrogram-based applications, such as instantaneous frequency estimation. To highlight the advantages of TFR enhancement through adaptive window length in the time domain, we apply the instantaneous frequency tracking method proposed in \cite{leclere2016multi} to both the spectrogram optimized with a single window and the spectrogram optimized with a time-varying window. The first row of Fig. \ref{fig:fig5} displays the tracked instantaneous frequencies for the tenth harmonic (dashed line). Dividing these tracked frequencies by 10 provides an estimate of the instantaneous frequency of the main harmonic. When a single window is used in the optimization, the mean square error of the estimated instantaneous frequency of the main harmonic is 7.05. In contrast, when a time-varying window length is employed, the mean square error is significantly reduced to 2.86, as indicated in the last row of Fig. \ref{fig:fig5}. This demonstrates the effectiveness of optimizing the window length in time for improved frequency tracking accuracy in vibration analysis.

\section{Applications: Task-Driven Optimization}
\label{sec:task_driven}

This section introduces the concept of task-driven optimization using the DSTFT, where the optimization aims to minimize a performance metric on a dataset relevant to a specific task. This contrasts with the earlier examples that focused on representation-driven optimization on individual signals based on TFR criteria. This section will illustrate this concept with two applications: joint optimization with frequency tracking and joint optimization with a neural network.

\subsection{Window Length Optimization for Frequency Tracking from a Vibration Signal}
\label{sec:freq_track}

\begin{figure}[b]
    \centering
    \begin{tikzpicture}
        \node[inner sep=0pt] (img1) {\includegraphics[width=5cm]{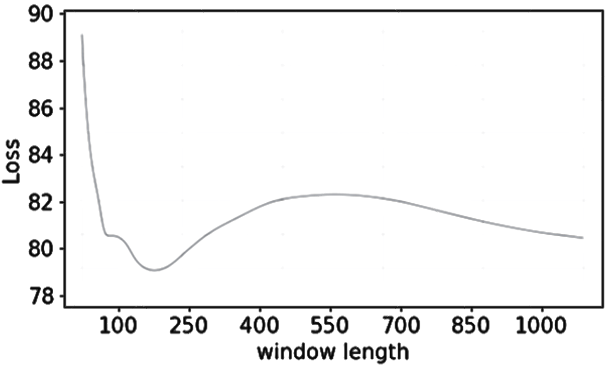}};
        \node[inner sep=0pt] (img2) at (img1.south) [below=0cm] {\includegraphics[width=5cm]{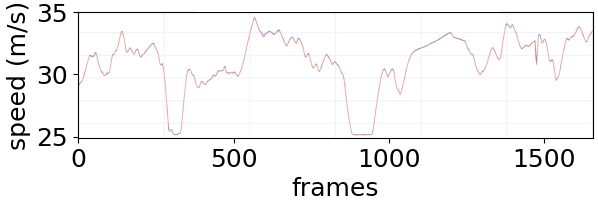}};
        \node[inner sep=0pt] (img3) at (img2.south) [below=0cm] {\includegraphics[width=5cm]{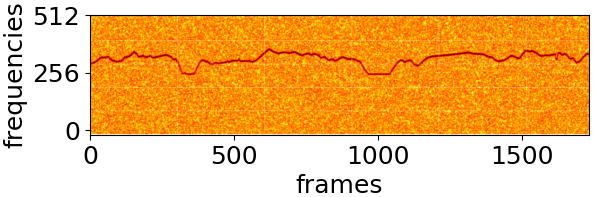}};
    \end{tikzpicture}
    \caption{Loss per window length (top), angular speed associated with optimal window length for a particular sample (middle), and the corresponding spectrogram (bottom).}
    \label{fig:fig6}
\end{figure}

In this experiment, the objective shifts from optimizing the window length for individual signals to determining a single, global window length that yields accurate frequency estimation on average across a dataset of signals. The training data for this task consists of $J$ synthetic variable-period sine waves corrupted by additive white noise. The goal is to find the window length $\winlength$ that minimizes the mean squared error between the estimated and true frequencies over the entire training dataset, as defined by the following loss function:
\begin{equation}
    \loss \circ (\hat{y}_j(\spec (\winlength)), \bar{y}_j) = \frac{1}{J}\sum_{j} \left\Vert\hat{y}_j ( \spec (\winlength))- \bar{y}_j\right\Vert^2,
    \label{eq::15}
\end{equation}
where $J$ is the number of signals in the training dataset, $\bar{y}_j$ is the ground truth frequency of the $j^{th}$ signal of the training data and $\hat{y}_j (S(\winlength))$ is a differentiable estimation of the frequency from the spectrogram of the $j^{th}$ signal computed with a window length $\winlength$. Note that the temporal positioning of the frames are fixed according to a predetermined hop-length $H$. For simplicity, we consider the frequency estimate for the $j^{th}$ signal at each time frame n as a weighted average of the frequency bins:
\begin{eqnarray}
\hat{y}_j (S(\theta)) = \left (\sum_\f  
\frac{|\spec_{\tap}^j (t_n,\f,\theta)|} 
{\sum_l | \spec_{\tap}^j (t_n,l,\winlength)|}  \f \Delta f \right )_n,
\label{eq::14}
\end{eqnarray}
where $\Delta f$ is the frequency resolution. 
We employ the gradient descent algorithm to minimize the loss function in Eq. \eqref{eq::15} with respect to $\winlength$. Upon convergence, the window length reaches a value that minimizes the loss. The first row of Fig. \ref{fig:fig6} shows the loss function as a function of the window length. The second row of Fig. \ref{fig:fig6} illustrates the resulting estimated angular speed on a sample from the simulated dataset, and the third row displays the spectrogram of that sample obtained using the optimized window length, which appears well-suited for the frequency tracking task. This simple simulation demonstrates the effectiveness of DSTFT-based backpropagation optimization for task-driven parameter learning.
This approach can be generalized to any signal processing task that can be formulated as the minimization of a loss function defined with respect to the spectrogram and some target data. Indeed, one can simply replace the standard spectrogram computation with the DSTFT-based spectrogram and then optimize the window parameters by minimizing the task-specific loss function using gradient descent based on the derived backpropagation formulas.

\subsection{Joint Optimization with a Neural Network}

This experiment investigates the advantage of optimizing the window length in DSTFT within a neural network training process. This approach contrasts with conventional methods where the STFT parameters are typically fixed before the spectrogram is used as input to the network \cite{badshah2017speech, gong2021ast, schluter2014improved, park2019specaugment, defossez2021hybrid}. Traditional methods often involve generating spectrograms with fixed parameters chosen through heuristics or discrete optimization techniques. By treating the DSTFT as a differentiable layer with a trainable real-valued parameter (the window length in this case), we can optimize this parameter alongside the network weights to minimize a task-specific loss function in an end-to-end manner. This allows the TFR to adapt to the requirements of the classification task through global optimization.

In this experiment, the DSTFT is integrated as the initial layer of a Convolutional Neural Network (CNN) designed for spoken digit classification using the Free Spoken Digit Dataset (FSDD), also known as Audio MNIST. This dataset comprises 3000 recordings of spoken digits from 6 speakers. We randomly partitioned the dataset into 80\% for training, 10\% for validation, and 10\% for testing. The primary goal is to jointly optimize the window length of the spectrogram generated by the DSTFT and the weights of the subsequent CNN layers. For a classification task with C=10 classes (digits 0-9), the optimization is performed by minimizing the cross-entropy loss between the CNN's predicted digit labels and the ground truth labels over the entire training dataset. For each sample j, the loss is defined as:
\begin{equation}
\label{eq:cross_entropy}
\mathcal{L}(y_j,\hat{y}_j) = - \sum_{c} y_j^c \log(\hat{y}_j^c),
\end{equation}
where $y_j^c$ is a one-hot encoded ground truth label (i.e.,  
$y_j^c=1$ if the true label for the $j^{th}$ sample is $c$,  
and 0 otherwise), and $\hat{y}_j = (\hat{y}_j^c)_{c=1,\dotsc,C}$ is the predicted probability of the $j^{th}$ input signal belonging to the $c^{th}$ digit class, as output by the CNN classifier. The prediction process can be viewed as a function where the input audio signal $x_j$ is first transformed into a magnitude spectrogram using the DSTFT with a learnable window length $\winlength$, and this spectrogram is then fed into the CNN classifier $F$ with learnable weights $w$: 
\begin{equation}
\label{eq:prediction}
\hat{y}_j = F_w \left( | \spec_\winlength (x_j) | \right).
\end{equation}
The optimization of both $\winlength$ and $w$ is performed using the Adam optimizer, a gradient-based algorithm that requires the computation of the partial derivatives of the loss function with respect to both sets of parameters ($\partial_\winlength \loss $ and $\partial_w \loss $), which can be derived using the chain rule and the differentiability of the DSTFT (as detailed in Sec. \ref{sec:4_grad}).

To evaluate the impact of the window size, a 2-layer CNN with 16 filters per layer was trained. Depthwise separable convolutions were used to balance computational efficiency and model accuracy. The network architecture also included ReLU activation functions for the hidden layers, dropout for regularization to prevent overfitting, and a final dense layer to produce the probability distribution over the ten digit classes. The entire system, from the input audio to the digit prediction, can be seen as a single neural network where the first layer is the DSTFT (with the window length as a trainable parameter), followed by the convolutional layers and the dense layer.
 Table \ref{tab:table1} shows the test loss of this CNN when trained with spectrograms computed using different fixed window lengths, highlighting the sensitivity of the performance to this parameter. Table \ref{tab:table3} presents the corresponding classification accuracies in percentage. As shown in Table \ref{tab:table3}, the classification accuracy varies significantly with different fixed window lengths, further emphasizing the importance of this parameter. Notably, when the window length is jointly optimized with the network weights using the DSTFT, the model achieves a higher test accuracy (80.7\%) compared to the best fixed window length case (79.7\% at window length 40). 
While the presented results demonstrate the effectiveness of the proposed approach, higher classification accuracies could be achieved on this task by employing larger and more complex neural network architectures with a greater number of parameters. The relatively small CNN used in this experiment (2 layers with 16 filters per layer) was intentionally chosen to highlight the benefit of optimizing the window length parameter.


Current standard practices in similar tasks often involve training separate neural networks for a range of fixed STFT window sizes and subsequently selecting the network that yields the best performance on a validation set. Our proposed method offers a more efficient alternative by jointly optimizing the window length and the network parameters in a single end-to-end training process.
To conclude, these two simple simulations demonstrate the effectiveness of our backpropagation procedure based on DSTFT. They illustrate a general window length tuning methodology applicable to any existing signal processing algorithm or neural network involving spectrograms: replace the standard spectrogram computation step with a DSTFT-based spectrogram, and then optimize the window length using gradient descent based on the derived backpropagation formulas. Differentiable STFT enables the use of gradient-based optimization for tuning its parameters in an end-to-end manner and optimizing the entire pipeline jointly.

\begin{table}[ht]
\centering
\caption{Training, validation and testing losses of NN with fixed and learnable window length.}
\label{tab:table1}
\begin{tabular}{|c|c|c|c|c|}
\hline 
approaches & window length &  train loss &  val loss &  test loss  \\ 
 \hline 
  STFT  & 10 &    0.37 &  1.15 &   1.04 \\
  STFT  & 20 &    0.03 &  0.44 &   0.32 \\
  STFT  & 30 &    0.01 &  0.24 &   0.27 \\
  STFT  & 40 &    0.01 &  0.26 &   0.23 \\
  STFT  & 50 &    0.01 &  0.49 &   0.29 \\
  DSTFT & $\winlength=34.9$ & 0.01 & 0.20 & 0.22 \\
  \hline 
\end{tabular}
\end{table}

\begin{table}[ht]
\centering
\caption{Training, validation and testing accuracy (in percentage) of CNN with fixed and learnable window length.}
\label{tab:table3}
\begin{tabular}{|c|c|c|c|c|}
\hline 
approaches & window length &  train loss &  val loss &  test loss  \\ 
 \hline 
    STFT & 10 & 68.0 & 61.3 & 62.0 \\
    STFT & 20 & 73.0 & 75.7 & 74.3 \\
    STFT & 30 & 79.3 & 80.3 & 79.7 \\
    STFT & 40 & 80.7 & 80.0 & 79.7 \\
    STFT & 50 & 80.7 & 76.7 & 79.0\\
    DSTFT & $\winlength=34.9$ & 80.3 & 82.3 & 80.7 \\
  \hline 
\end{tabular}
\end{table}


\section{Conclusion}
\label{sec:6_ccl}
We have introduced a differentiable formulation of the short-time Fourier transform that enables gradient-based optimization of window lengths and temporal positions. This approach offers advantages such as adaptive and automatic tuning of parameters for STFT-based TF representations like spectrograms. The significance of this contribution lies also in its potential applications within machine learning, where differentiable models are essential for efficient optimization algorithms.

\bibliographystyle{IEEEtran}
\bibliography{refs}

\end{document}